*Genome Analysis*

# TreeGrafter: phylogenetic tree-based annotation of proteins with Gene Ontology terms and other annotations

Haiming Tang[1], Robert D Finn[2], Paul D Thomas[1, *]

[1]Division of Bioinformatics, Department of Preventive Medicine, University of Southern California, Los Angeles, CA 90033, USA.

[2]European Molecular Biology Laboratory, European Bioinformatics Institute (EMBL-EBI), Wellcome Trust Genome Campus, Hinxton, Cambridge CB10 1SD, UK.

Associate Editor: XXXXXXX

**ABSTRACT**

**Summary:** TreeGrafter is a new software tool for annotating protein sequences using annotated phylogenetic trees. Currently, the tool provides annotations to Gene Ontology terms, and PANTHER protein class, family and subfamily. The approach is generalizable to any annotations that have been made to internal nodes of a reference phylogenetic tree. TreeGrafter takes each input query protein sequence, finds the best matching homologous family in a library of pre-calculated, pre-annotated gene trees, and then grafts it to the best location in the tree. It then annotates the sequence by propagating annotations from its ancestral nodes in the reference tree. We show that TreeGrafter outperforms subfamily HMM scoring for correctly assigning subfamily membership, and that it produces highly specific annotations of GO terms based on annotated reference phylogenetic trees. This method will be further integrated into InterProScan, enabling an even broader user community.

**Availability:** TreeGrafter is freely available on the web at https://github.com/haimingt/TreeGrafting.

## 1 INTRODUCTION

The growing rate of discovery of new protein sequences continues to increase the demand for automated computational methods for functionally annotating these sequences. The Gene Ontology (GO) is by far the most highly used, computationally accessible representation of gene and protein function (Ashburner, et al., 2000; The Gene Ontology Consortium, 2017). Several methods have been developed to infer GO annotations for experimentally uncharacterized protein sequences. Blast2GO find homologs of input sequences using BLAST, extracts existing GO annotations for obtained hits, and finally assigns GO terms for query sequences using an annotation rule (Conesa, et al., 2005). InterPro2GO (Burge, et al., 2012) associates GO terms with InterPro entries, and propagates GO terms tosequences based on matching InterPro entries (Mitchell, et al., 2015). PANTHER (Mi, et al., 2017) classifies sequences using two types of HMM: family HMMs (that recognize members of a large family tree) and subfamily HMMs (that recognize members of a sub-family within the family tree) and similarly annotates the query sequence with the GO annotations of the matching HMMs.

Over the past few years, the GO Consortium has made substantial progress in annotating gene trees with GO terms using the Phylogenetic Annotation and INference Tool (PAINT) (Gaudet, et al., 2011). This tool helps curators to make precise assertions as to when functions were gained and lost during evolution, and record the evidence (experimentally supported GO annotations and phylogenetic relationships) for those assertions. In this way, PAINT makes it possible to handle conservation and divergence for each function on a case-by-case basis, decreasing false positive and false negative function prediction rates (Gaudet, et al., 2011). To date over 4500 families have been annotated with thousands of "functional evolution events": gain and loss of gene function at specific nodes in the evolutionary trees. PAINT has been used to annotate protein sequences from the 104 genomes in these reference trees, but until now there has been no way to apply these annotations to the millions of sequences uncovered by other sequencing projects, both whole genome and metagenome.

Here we present a new tool, TreeGrafter, which extends the tree-based annotation inference model to sequences that are not in the annotated reference tree. TreeGrafter grafts a query sequence onto the reference phylogenetic tree. Like any other sequence in the tree, the query sequence will inherit annotations (including function annotations, family label annotations, etc.) from its annotated ancestral nodes in the tree.

## 2 METHODS

### 2.1 Trees, alignments and GO annotated gene trees from PAINT

Sequence alignments and phylogenetic trees, and annotated subfamily nodes were obtained from the PANTHER database, version 12.0 (Mi et al. 2017). GO annotations for nodes in the gene trees (Gaudet, et al., 2011) were obtained from the Gene Ontology github repository. Currently, approximately one third (4650/14710) of the PANTHER families have been curated with GO annotations.

## 2.2 TreeGrafter algorithm: grafting and inheritance of annotations

Our grafting approach is similar to the TreeFam "orthology on the fly" tool (Schreiber, et al., 2014), but differs in how alignments are computed and how the graft point is determined. Pairwise alignments between a sequence and profile HMM (generated using hmmscan from HMMER3) are used to add the query sequence to a precomputed multiple alignment (produced by aligning the tree sequences to the profile HMM), rather than MAFFT, which was found to be the rate-limiting step in most cases.

The grafting of a new sequence onto the reference tree proceeds in three steps: (i) the query is scored against the PANTHER HMM library (family HMMs only) to find a best matching reference phylogenetic tree (Mi et al. 2017) and to obtain an alignment to the family HMM; (ii) the alignment to the HMM is used to add the query to the pre-calculated multiple sequence alignment; (iii), RAxML version 8 (Stamatakis, 2014) (using parsimony mode for efficiency) is used to graft the query sequence onto the reference tree with the extended alignment that contains all sequences (reference + query) as input. If multiple, equally parsimonious graft points are found, the last common ancestor in the tree is used as the consensus graft point. From the graft point, ancestral annotations are traced through a propagation process from the root of the reference tree to the position of the query sequence. In this process, ancestral gain of function annotations are inherited, and ancestral loss of function annotations prevent propagation (Figure 1).

## 3 TESTING AND RESULTS

### 3.1 Accuracy of tree grafting

We tested TreeGrafter performance for grafting to the correct tree position, using eight complete proteomes from different taxa: *Equus caballus* (horse), *Anolis carolinensis* (green anole lizard), *Anopheles gambiae* (mosquito), *Populus trichocarpa* (poplar), *Cryptococcus neoformans JEC21* (fungus), *Methanosarcina acetivorans* (archaeon), *Salmonella typhimurium LT2* (bacterium). For each sequence, we first remove it from the corresponding PANTHER phylogenetic tree and multiple sequence alignments, and then graft the input sequence back to the reduced tree using TreeGrafter. The sequence inherits a subfamily label based on its position in the tree.

TreeGrafter outperformed subfamily HMM scoring (the standard used by PANTHER and in InterProScan for nearly 20 years) for assigning sequences to the proper subfamily (Supplemental Table 1). This test was particularly stringent as we removed the validation sequences from the reference trees (and alignment), but not from the alignments used to train the subfamily HMMs. Using HMMER3 substantially increases speed (Supplemental Figure 1) and also marginally increases performance on our subfamily classification benchmark (Supplemental Table 1).

### 3.2 Comparing GO annotations from TreeGrafter with InterPro2GO

Interpro2GO (Burge, et al., 2012) is the state-of-art and one of the most widely used tools for protein sequence annotation. InterPro signatures (primarily HMMs, including PANTHER) have been annotated with GO terms by expert curation. We compared the GO annotations from TreeGrafter and InterPro2GO for each protein sequence of the eight species (Supplemental Table 2). Overall, we find that for annotated proteins, TreeGrafter infers a larger number of GO annotations than InterPro2GO. When GO terms from the two methods are related in the GO hierarchy (and hence comparable), TreeGrafter annotations tend to be more specific. However, GO annotations from TreeGrafter do not completely overlap with InterPro2GO, and do not currently cover as many proteins, demonstrating the complementarity of the approaches. TreeGrafter will be corporated into InterProScan in the near future, and the number of proteins annotated by TreeGrafter will continue to increase as the GO Phylogenetic Annotation project proceeds.

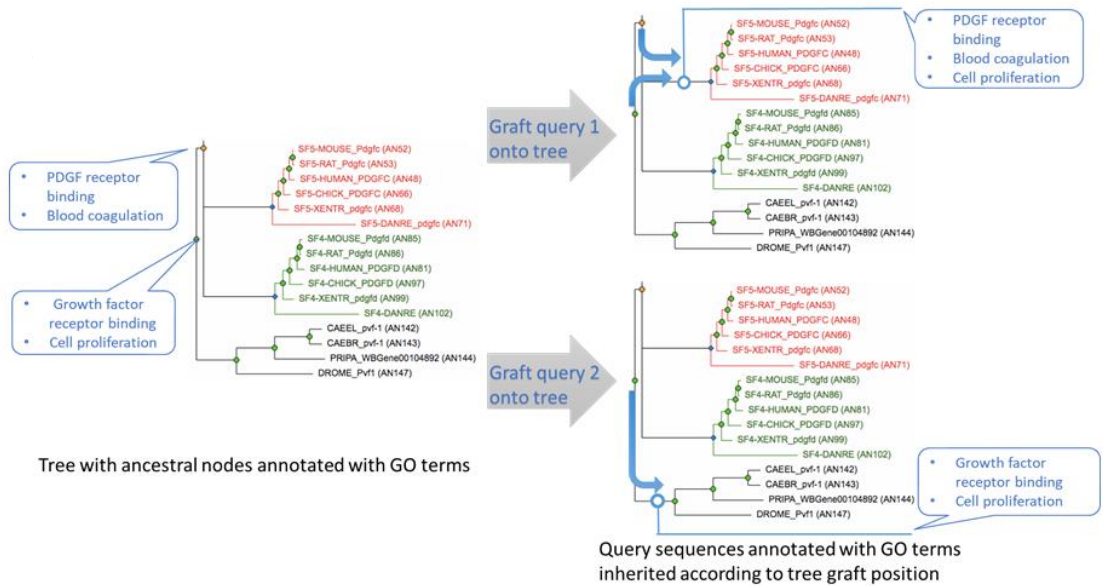

*Figure 1 - TreeGrafter annotates each sequence based on where it is grafted onto an annotated reference tree. Given the same tree with pre-annotated ancestral gene nodes (left panel), each query sequence is grafted onto the tree. For the graft position position of query 2 (bottom, blue open circle), there is only one annotated ancestral node and only the annotations from this one node are inherited by query 2.*

## 4 IMPLEMENTATION

TreeGrafter is implemented in Perl as a standalone command line tool. The code is available at https://github.com/haimingt/TreeGrafting.

## 5 CONCLUSIONS

TreeGrafter is an efficient tool for annotating large protein sets, such as those derived from whole genome or metagenome sequencing. It currently annotates subfamily labels from PANTHER, and Gene Ontology terms from the GO Phylogenetic Annotation project, but can be generalized to any annotations made to ancestral nodes in a reference gene tree.

## ACKNOWLEDGEMENTS

We thank Dr. Huaiyu Mi for help with the PANTHER and PAINT data. This work was supported by the National Science Foundation (grant number 1458808).